\documentclass[conference]{IEEEtran}
\IEEEoverridecommandlockouts
\usepackage{cite}
\usepackage{amsmath,amssymb,amsfonts}
\usepackage{float}
\usepackage{wrapfig}
\usepackage{graphicx}
\usepackage{subcaption} % for subfigures
\usepackage{textcomp}
\usepackage{xcolor}
\usepackage{float}
\usepackage{xspace}
\usepackage{booktabs}
\usepackage{amsmath}
\usepackage{tikz}      % for creating diagrams
\usepackage[utf8]{inputenc}
\usepackage{color}
\usepackage{tikz}
\usepackage{mathtools}
\usetikzlibrary{3d}
\usepackage{xcolor}
\usepackage{soul}
\usepackage{comment}
\usepackage{hyperref}
\usepackage{longtable}
\usepackage{multirow}
\usepackage{lineno}
\usepackage{blochsphere}  % for bloch sphere representation

\usepackage{algorithm}
\usepackage{bm}      % for bold vectors (|\psi\rangle)
\usepackage{braket}
\usepackage{algpseudocode}
\usepackage{threeparttable}
\hypersetup{
    colorlinks=true, % Enable colored links
    linkcolor=blue,  % Color for internal links (e.g., table of contents, cross-references)
    citecolor=blue, % Color for citations
    urlcolor=red     % Color for URLs
}
\def\BibTeX{{\rm B\kern-.05em{\sc i\kern-.025em b}\kern-.08em
    T\kern-.1667em\lower.7ex\hbox{E}\kern-.125emX}}
\begin{document}

\title{Benchmarking of GPU-optimized Quantum-Inspired Evolutionary Optimization Algorithm using Functional Analysis}

\author{

\IEEEauthorblockN{Kandula Eswara Sai Kumar}
 \IEEEauthorblockA{\textit{Principal Optimization Scientist}
 } 
\and 

\IEEEauthorblockN{Supreeth B S}
 \IEEEauthorblockA{\textit{Quantum Algorithm Researcher}
 }
 \and
\IEEEauthorblockN{Rajas Dalvi}
 \IEEEauthorblockA{\textit{Quantum Optimization Researcher} 
 }
 \and
\IEEEauthorblockN{Aman Mittal}
 \IEEEauthorblockA{\textit{HPC Researcher} 
 }
 \and

\IEEEauthorblockN{Aakif Akhtar \hspace{1.5cm} Ferdin Don Bosco \hspace{1.5cm} Rut Lineswala \hspace{1.5cm}  Abhishek Chopra}
\IEEEauthorblockA{\textit{ Quantum Algorithm Developer \quad Senior Computational Scientist\quad Chief Technology Officer \quad CEO \& Chief Scientific Officer \qquad }
\\
\\
%\begin{center}
\parbox{\linewidth}{\centering
BosonQ Psi (BQP), New York, USA \\
eswara.sai@bosonqpsi.com \quad abhishek.chopra@bosonqpsi.com
}
%\end{center}
}
%}
}
\maketitle

\begin{abstract}
This article presents a comparative analysis of GPU-parallelized implementations of the quantum-inspired evolutionary optimization (QIEO) approach and one of the well-known classical metaheuristic techniques - the genetic algorithm (GA). 
The study assesses the performance of both algorithms on highly non-linear, non-convex, and non-separable function optimization problems, viz., Ackley, Rosenbrock, and Rastrigin, that are representative of the complex real-world optimization problems. 
The performance of these algorithms is checked by varying the population sizes by keeping all other parameters constant and comparing the fitness value it reached along with the number of function evaluations they required for convergence.
The results demonstrate that QIEO performs better for these functions than GA, by achieving the target fitness with fewer function evaluations and significantly reducing the total optimization time— approximately three times for the Ackley function and four times for the Rosenbrock and Rastrigin functions. 
Furthermore, QIEO exhibits greater consistency across trials, with a steady convergence rate that leads to a more uniform number of function evaluations, highlighting its reliability in solving challenging optimization problems. 
The findings indicate that QIEO is a promising alternative to GA for these kind of functions.
\end{abstract}

\begin{IEEEkeywords}
Performance Analysis, Optimization, Accuracy, Quantum-inspired
\end{IEEEkeywords}

\section{Introduction}
Traditional optimization methods, such as gradient descent, Newton-Raphson method, etc., are powerful but have notable drawbacks. Firstly, they are local in scope and often fail to find high-quality solutions in complex, non-convex landscapes. Secondly, these algorithms rely on gradient information throughout the search process, necessitating that the design space be continuous and smooth, which is often not the case in the real world \cite{Rao2009}. Additionally, they can be computationally expensive, especially in high-dimensional spaces, due to the need for frequent gradient evaluations. When the design space is discrete, researchers have employed enumeration algorithms; however, as the number of design variables grows, the performance of these algorithms rapidly deteriorates as they suffer from the curse of dimensionality \cite{bellman1954dynamic}

The utilization of meta-heuristics, such as Genetic Algorithm (GA) \cite{Holland1992, Goldberg1988}, Particle Swarm Optimization \cite{Kennedy}, and Ant Colony Optimization \cite{Dorigo1996}, have become powerful tools for tackling such challenges. 
Inspired by natural processes, these algorithms efficiently find near-optimal solutions for a wide range of optimization problems in reasonable time frames \cite{doi:10.1080/18756891.2015.1046324, Degertekin2016}. 
These algorithms have demonstrated considerable potential in solving optimization problems where exact algorithms are impractical. 
These methods offer efficiency and scalability for multiple objectives \cite{Deb2002}, and they can be readily adapted to various problem domains, including combinatorial, discrete, and continuous optimization \cite{Blum2003}. 
They exhibit robustness to imperfect objective functions, which are often discontinuous, non-differentiable, and erratic \cite{Michalewicz2004}, and can be implemented in distributed and parallel computing infrastructures \cite{Talbi2009}. 
However, challenges arise in these meta-heuristic algorithms from their dependence on extensive hyper-parameter tuning, which complicates the development of a general-purpose optimization method. 
Although these algorithms are intended to be general-purpose, their performance can be inconsistent, unreliable, and time-consuming \cite{KIM2012564,zhou2023}. 
This inconsistency arises from the need to adjust numerous hyper-parameters and the inherent randomness in their processes, making it difficult to achieve predictable results in any given trial \cite{6121807}. 

In engineering optimization, the problem's dimensionality can reach as high as on the order of $10^3$ or $10^4$. This might necessitate very large population sizes, thereby increasing the total number of function evaluations, with each individual requiring a Finite Element Method (FEM) evaluation over multiple iterations that can span from several hours to days. This can prove to be very costly. 
For instance, topology optimization \cite{bendsoe2013topology,Kumar2020}, shape optimization \cite{8727653}, flight trajectory optimization \cite{Dai2009}, aircraft routing optimization \cite{ARNOLD201932}, supply chain optimization \cite{SCHULZ20051305}, energy storage optimization \cite{6461989}, job shop scheduling optimization \cite{HUANG2017642} and traveling salesman problem \cite{JIANG2020112867}, all of these different optimization problems require a higher computational time for each function evaluation. 
Therefore, it is crucial for an algorithm to achieve the desired fitness with fewer function evaluations consistently since the evaluations are expensive. 

All of these factors motivated us to develop and benchmark a quantum meta-heuristic algorithm, namely, Quantum-Inspired Evolutionary Optimization (QIEO). It is characterized by the independent evolutionary strategy \cite{870809, 10.1145/2598394.2598437} and reliance on only one parameter-dependent operator, which offers the potential to mitigate the discussed limitations of conventional meta-heuristics. 
Inspired by the probabilistic principles of quantum physics, these algorithms have demonstrated significant potential in optimization.

In this article, we assess the performance of the GPU-optimized QIEO algorithm by comparing it with the GPU-optimized GA using benchmark functions. 
The evaluation focuses on two key metrics: the number of function evaluations and the consistency in achieving the desired accuracy and convergence rate across multiple trials. 
Benchmark functions play a crucial role in understanding the algorithm’s effectiveness in solving real-world optimization problems. 
For this purpose, we selected complex, multi-modal, multidimensional, and non-separable functions from the CEC2014 benchmark suite \cite{liang2013problem}, which closely reflect the challenges encountered in practical optimization scenarios.

\section{Methodology}
\subsection{Background on quantum-inspired evolutionary algorithm}

QIEO algorithm aims to improve optimization processes by leveraging quantum mechanical principles. 
These algorithms are designed to execute on classical high-performance computers while being adaptable for execution on quantum hardware \cite{ACAMPORA2021542}. 
Their objective is to explore solution spaces more effectively, potentially outperforming traditional optimization strategies, particularly for specific optimization problems \cite{Zhang2011}.

The principles of quantum mechanics, such as qubits, quantum superposition, quantum gates, and quantum measurement, are mainly used for developing QIEO algorithms. The classical computer uses bits, which can be either 0 or 1. On the other hand, a qubit (the fundamental unit of information in a quantum computer) can exist in a superposition of both states. However, a qubit collapses upon measurement into a classical state, either 0 or 1. 
In the algorithm, it constitutes a quantum gene, which can be represented as follows:
\begin{equation}
\begin{bmatrix}
\alpha \\
\beta
\end{bmatrix}
\end{equation}
The state of the quantum gene can be mathematically described as:
\begin{equation}
|\psi\rangle = \alpha |0\rangle + \beta |1\rangle
\label{eq: qubit definition}
\end{equation}
Here $\alpha$ and $\beta$ are complex numbers that specify the probability amplitudes associated with the classical states $\ket{0}$ and $\ket{1}$, respectively.  Their squared magnitudes, $|\alpha|^2$ and $|\beta|^2$, denote the probability of observing the qubit in the states $\ket{0}$ and $\ket{1}$, respectively, upon measurement. When m qubits are considered, they can together exist in a superposition of $2^m$ classical states. These m-qubits form a quantum individual, which is depicted as follows: 
\begin{equation}
\textbf{\textit{q}}_{j}=
    \begin{bmatrix}
        \alpha_1 &\alpha_2& \dots& \alpha_m \\
        \beta_1 &\beta_2 & \dots & \beta_m
    \end{bmatrix}
    \label{equation:quantum chromosome}
\end{equation}
The state of the quantum individual is given by:
\begin{equation}
\ket{\psi}^{\otimes m} = \sum_{\mathbf{x} \in \{0,1\}^m}  p_\mathbf{x} \ket{\mathbf{x}},
\end{equation}
Here, $\mathbf{x}$ represents a classical state with $m$ bits, and $ p_\mathbf{x}$   represents its probability amplitude.
For example, a system of three qubits can exist in a superposition of eight classical states whose quantum individual is as follows:
\begin{equation}
\textbf{\textit{q}}_{j}=
    \begin{bmatrix}
        \alpha_1 &\alpha_2&\alpha_3 \\
        \beta_1 &\beta_2&\beta_3
    \end{bmatrix}
\end{equation}
The state of the quantum individual is given by: 
\begin{equation}
\begin{aligned}
\left| \psi \right\rangle^{\otimes 3} &= (\alpha_1\alpha_2\alpha_3)\left| 000 \right\rangle\\
&+ (\alpha_1\alpha_2\beta_3) \left| 001 \right\rangle + (\alpha_1\beta_2\alpha_3) \left| 010 \right\rangle \\
&+ (\alpha_1\beta_2\beta_3) \left| 011 \right\rangle + (\beta_1\alpha_2\alpha_3) \left| 100 \right\rangle \\
&+ (\beta_1\alpha_2\beta_3) \left| 101 \right\rangle + (\beta_1\beta_2\alpha_3) \left| 110 \right\rangle \\
&+ (\beta_1\beta_2\beta_3)\left| 111 \right\rangle \\
\end{aligned}
\end{equation}
where $p_{000}=\alpha_1\alpha_2\alpha_3$, $p_{001}=\alpha_1\alpha_2\beta_3$ and so on. A group of $n$ such quantum individuals form a quantum population \boldmath $Q = \{q_1, q_2, \dots, q_n\}$\unboldmath. Upon measurement, a classical population \boldmath $P = \{  p_1, p_2, \dots, p_n \}$ \unboldmath is created. Having understood this, we will describe the quantum-inspired evolutionary algorithm. 
\begin{algorithm}[H]
\caption{Quantum-Inspired Evolutionary Optimization Algorithm}
\label{alg:QIEO}
\begin{algorithmic}[1]
    \State $t \leftarrow 0$
    \State Initialize quantum parallelism to form \textbf{\textit{Q(t)}}.
    \State Obtain the population \textbf{\textit{P(t)}} by observing \textbf{\textit{Q(t)}}.
    \State Perform classical information processing by evaluating \textbf{\textit{P(t)}}, storing the best solution $b$, and obtaining the $\theta$ for $R_Y(t)$ gate.
    \While{(not termination condition)}
        \State $t \leftarrow t+1$
        \State Perform quantum information processing to exploit the 
        \Statex \hspace{\algorithmicindent}quantum parallelism: Evolve \boldmath$Q(t-1)$\unboldmath  using a $U$-
        \Statex\hspace{\algorithmicindent}gate to obtain \boldmath$Q(t)$\unboldmath: $R_Y(t-1) \cdot$ \boldmath$Q(t-1)$\unboldmath =  \boldmath$Q(t)$\unboldmath.
        \State Obtain the population \boldmath$P(t)$ \unboldmath  by observing \boldmath$Q(t)$\unboldmath.
        \State Evaluate \boldmath$P(t)$\unboldmath, store the best solution $b$ till the  $t^{th}$ 
        \Statex \hspace{\algorithmicindent}iteration, and obtain the $\theta$s for $R_Y(t)$ gate.
    \EndWhile
\end{algorithmic}
\end{algorithm}
\hspace{-0.4cm}The details of the QIEO algorithm are as follows: \\
\textbf{Step 1}: The iteration counter t is initialized to 0.\\ 
\textbf{Step 2 -- Initialization of quantum parallelism:}
A population of quantum individuals \boldmath  $Q(t) = \{q_1^t, q_2^t, \dots, q_n^t\}$ \unboldmath is created such that $\alpha$s and $\beta$s of each 
$\textbf{\textit{q}}_{j}^{t}$ (in Eq.(\ref{equation:quantum chromosome})) are set to $1$ and $0$, respectively.
The state of each individual is given by: 
\begin{equation}
\left| \psi \right\rangle^{\otimes m}= |0_1 0_2 \dots 0_m\rangle
\end{equation}
The Hadamard gate is then applied which evolves the  $\alpha$s and $\beta$s of each $\textbf{\textit{q}}_{j}^{t}$ to 
$\frac{1}{\sqrt{2}}$, creating the quantum state for the individual: 
\begin{equation}
H^{\otimes m}\ket{0^m} =  \sum_{\mathbf{x} \in \{0,1\}^m} \frac{1}{\sqrt{2^m}}\ket{\mathbf{x}},     
\end{equation}

\textbf{Step 3 -- Measurement:}
A classical population \boldmath $P(t) = \{  p_1^t, p_2^t, \dots, p_n^t \}$ \unboldmath is obtained by making a measurement on \boldmath $Q(t)$\unboldmath, where each $\textbf{\textit{p}}_{j}^{t}$ is a binary string of length $m$. 
The measurement operation on the $i^{th}$ qubit of the $j^{th}$ individual is as follows: it randomly generates a number $r$ between 0 and 1. It then compares this random number to the qubit's $|\alpha|^2$. 
If the random number is greater than $|\alpha|^2$, the corresponding classical bit $x$ is set to 1; otherwise, it's set to 0. This process is carried out on all the individuals' qubits, effectively simulating the collapse of the quantum state into a classical state based on the probabilistic measurement. 
\textbf{Step 4 -- Classical information processing:}
The obtained classical population is mapped to points within the search space. Their fitness is then evaluated through the objective function, and the fittest solution among \boldmath $P(t)$ \unboldmath is stored as \boldmath $b$\unboldmath. The $R_Y(t)$ gate parameter $\theta$ corresponding to the $i^{th}$ qubit of $\textbf{\textit{q}}_{j}^{t}$ is obtained by comparing the $i^{th}$ bit of \boldmath $b$ \unboldmath and the $i^{th}$ bit of $\textbf{\textit{p}}_{j}^{t}$ in \boldmath $P(t)$\unboldmath. The $\theta$s for all the qubits are determined such that when $\textbf{\textit{q}}_{j}^{t}$ is evolved through the $R_Y$ gate, it has a slightly higher likelihood of generating individuals similar to that of \boldmath $b$ \unboldmath than before. This process is carried out to generate $\theta$s for all the quantum individuals.
\begin{algorithm}[H]
\caption{Classical information processing to obtain $\theta$, where ${i_b}$ is the ith bit of the fittest individual.}
\begin{algorithmic}[1]
\For{\boldmath $j$ \unboldmath in \emph{P(t)} } 
    \For{ $i$ in \emph{${p_j}'s $}bits}
        \If{$i = 0$ and $i_b = 0$}
        \State $\theta_{ji} \gets 0$
        \ElsIf{$i = 0$ and $i_b = 1$}
        \State $\theta_{ji} \gets + \Delta \theta$
        \ElsIf{$i = 1$ and $i_b = 0$}
        \State $\theta_{ji} \gets - \Delta \theta$
        \ElsIf{$i = 1$ and $i_b = 1$}
        \State $\theta_{ji} \gets 0$
        \EndIf
    \EndFor
\EndFor
\end{algorithmic}
\label{algo:classical_information_processing}
\end{algorithm}

\textbf{Step 5 and 6:} We enter the while loop and update the iteration counter to t = t+1. 

\textbf{Step 7 -- Quantum information processing:} 
The $\theta$s obtained for each qubit in each individual from classical information processing are used to construct the corresponding $R_Y(t-1)$ gate. The gate evolves the state of the qubit in $\textbf{\textit{q}}_{j}^{t-1}$ to a new state, and its action is shown as follows: 
\begin{equation}
\begin{bmatrix}
\alpha ^ {t} \\
\beta ^ {t}\\
\end{bmatrix} = \begin{bmatrix}
\cos(\theta) & -\sin(\theta) \\
\sin(\theta) &  \cos(\theta)
\end{bmatrix}
\begin{bmatrix}
\alpha ^ {t-1} \\
\beta ^ {t-1}\\
\end{bmatrix}
\end{equation} 
where $\begin{bmatrix}
\alpha ^ {t-1} \\
\beta ^ {t-1}\\
\end{bmatrix}$ and 
$\begin{bmatrix}
\alpha ^ {t} \\
\beta ^ {t}\\
\end{bmatrix}$ are the amplitudes of the qubit before and after the evolution, respectively, and the square matrix is the $R_Y$ gate. 
This process is repeated for all the quantum individuals' qubits to generate the quantum population \boldmath$Q(t)$\unboldmath.
Now that we have obtained the new quantum population \textbf{\textit{Q(t)}}, obtain \textbf{\textit{P(t)}} using \textbf{\textit{Q(t)}} as described in step 3 and then follow the procedure in step 4 with a minor change that the fittest string among \textbf{\textit{P(t=0,1,2...t)}} is stored as \textbf{\textit{b}}.  This is continued till the termination criterion is satisfied. 

\subsection{Background on Genetic algorithm}
For the comparative analysis, this article implemented a generation model genetic algorithm  that is presented as follows \cite{Eshelman1990TheCA}:  
\begin{algorithm}[H]
\caption{Generational Model Genetic Algorithm}
\label{algo:genetic_algorithm}
\begin{algorithmic}[1]
    \State $t \gets 1$
    \State Initialize random population $P(t)$
    \State  Evaluate $P(t)$
    \While{(not termination condition)}
    \State Binary tournament selection on $P(t)$ to create $S(t)$
    \State Crossover on $S(t)$ to create $C(t)$
    \State Mutation on $C(t)$ to create $M(t)$
    \State Evaluate $M(t)$
    \State Survival of the fittest among $(P(t), M(t))$ to create 
    \Statex\hspace{\algorithmicindent} $P(t+1)$.
    \State $t \gets t+1$
    \EndWhile
\end{algorithmic}
\end{algorithm}
The iteration counter begins at 1, followed by the initialization of a random population of size \( n \). These individuals are then mapped to points in the design variable space and evaluated.

Upon entering the while loop:

\begin{itemize}
    \item Binary tournament selection is performed on the \( n \) individuals of \( P(t) \) to form a mating pool of n individuals denoted as \( S(t) \).
    \item The individuals of \( S(t) \) are recombined, resulting in \( n \) offspring individuals, \( C(t) \).
    \item Subsequently, mutation is executed on \( C(t) \) to generate \( n \) individuals in \( M(t) \).
    \item \( M(t) \) is then mapped to points in the design variable space and evaluated.
\end{itemize}

Finally, the fittest individuals from both \( P(t) \) and \( M(t) \) are selected to proceed to the next iteration.

\subsection{Benchmarking on function optimization}\label{sec:function_opt}
Function optimization is crucial in assessing the effectiveness of newly developed optimization techniques. Test functions, also referred to as artificial landscapes, serve as standardized benchmarks to evaluate algorithm performance under controlled conditions. These functions are categorized based on various factors, such as the type of optimization problem they represent, their complexity, and their specific characteristics. This categorization aids in understanding how different algorithms perform across different issues and can provide insights into their strengths and weaknesses. Common categories of standard functions include: 
\begin{itemize}
    \item 
Unimodal Functions: These functions have a single optimal solution, making them relatively more straightforward to optimize than multimodal functions.
\item
Multimodal Functions: Unlike unimodal functions, multimodal functions have multiple local optima, posing a more significant challenge to optimization algorithms to find the global optimum.
\item
Separable Functions: These functions can be decomposed into independent sub-functions, allowing optimization algorithms to potentially optimize each sub-function separately.
\item
Non-Separable Functions: Unlike separable functions, non-separable functions have interdependencies between variables, making them more challenging to optimize.
\end{itemize}
The following three multi-modal (or non-convex) functions are selected for QIEO benchmarking :
\begin{enumerate}
    \item Ackley: The Ackley function is challenging for optimization algorithms due to its complex landscape, which includes many local minima and a narrow global minimum surrounded by a nearly flat region. This makes it difficult for algorithms to distinguish between local and global optima. The flat regions can cause slow convergence, while the sharp, isolated global minimum requires high precision for an algorithm to reach it. Additionally, the oscillatory nature of the function's surface complicates gradient-based methods. Together, these factors pose significant hurdles for optimization techniques and it is defined as: 
    \begin{equation*}
    \begin{split}
        f(\mathbf{x}) = & -20 \exp\left(-0.2 \sqrt{\frac{1}{n} \sum_{i=1}^{n}x_i^2}\right) \\
        & - \exp\left(\frac{1}{n} \sum_{i=1}^{n}\cos(2\pi x_i)\right) + 20 + \exp \left(1\right)
    \end{split}
    \end{equation*}
\item Rosenbrock: The Rosenbrock function is challenging due to its extremely narrow ridge, which makes optimization difficult. The ridge follows a parabolic curve with a sharp peak at its tip. Algorithms that struggle to identify promising search directions often perform poorly on this problem. This characteristic makes Rosenbrock a tough test for many optimization algorithms. The function is defined as follows:
\[ f(\mathbf{x}) = \sum_{i=1}^{n-1} \left[ 100 (x_{i+1} - x_i^2)^2 + (x_i - 1)^2 \right]
\]
\item Rastrigin: Rastrigin's function is notoriously challenging for optimization algorithms due to its vast search space complexity and numerous local minima. The function's surface is shaped by external parameters $A$ and $c$, which influence the amplitude and frequency of modulation, respectively. With $A = 10$ and $c = 2\pi$, the modulation dominates the selected domain. This function is highly multimodal, with local minima arranged in a rectangular grid of size $1$. As the distance from the global minimum increases, the fitness values of the local minima grow larger, and the function definition is as follows:
    \begin{equation*}
        f(\mathbf{x}) = 10n + \sum_{i=1}^{n}(x_i^2 - 10\cos(2\pi x_i))
    \end{equation*}
\end{enumerate}
The above three functions are highly nonlinear (increase the function complexity) and non-convex (increase the design space complexity). The global minimum of three functions is zero, and converging to the global minimum is a tedious task with gradient-based approaches as they converge to local minima with the help of search direction. The search direction for the non-linear and non-convex functions is highly dependent on the point where it is and involves more computation. Most engineering optimization problems in the real world have non-linear, multimodal functions as objective functions. By testing optimization algorithms on these standard functions of different categories, researchers can gain valuable insights into the performance characteristics of different algorithms and make informed conclusions regarding their applicability to real-world problems \cite{liang2013problem}.

\section{Results and discussions}
This section presents the results obtained by testing the GPU-optimized QIEO and GA on function optimization, with the functions as described in Section \ref{sec:function_opt}. GA and QIEO algorithms employed in these simulations are developed using CUDA C++, leveraging the computational power of an NVIDIA A100 GPU card. The Center for Computational Research at the University at Buffalo supports these computational facilities  \cite{ubuffsite}. Both algorithms use the same encoding of the design space and apply identical convergence criteria. The convergence criteria used for GA and QIEO are as follows:
\begin{itemize}
    \item Maximum number of generations: The algorithm terminates when it reaches 3000 generations, regardless of other factors.
    \item Fitness value tolerance: The algorithm stops when the change in fitness value reaches a tolerance, $1e^{-8}$, indicating that the solution is sufficiently optimized. 
\end{itemize}

\begin{figure*}[hbt]
    \centering
    \begin{subfigure}{0.47\textwidth}
        \centering
        \includegraphics[width=0.9\linewidth]{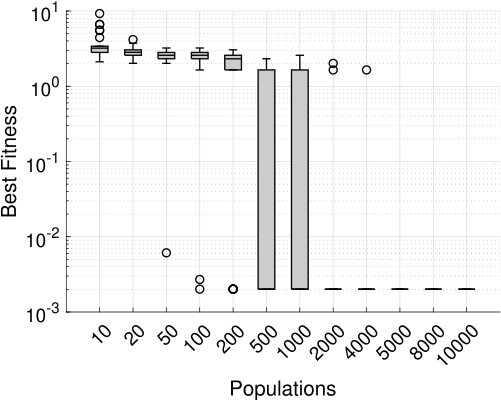}
        \caption{GA}
        \label{fig:AckleyPopvsBestFitGA}
    \end{subfigure}
    \hfill
    \begin{subfigure}{0.47\textwidth}
        \centering
        \includegraphics[width=0.9\linewidth]{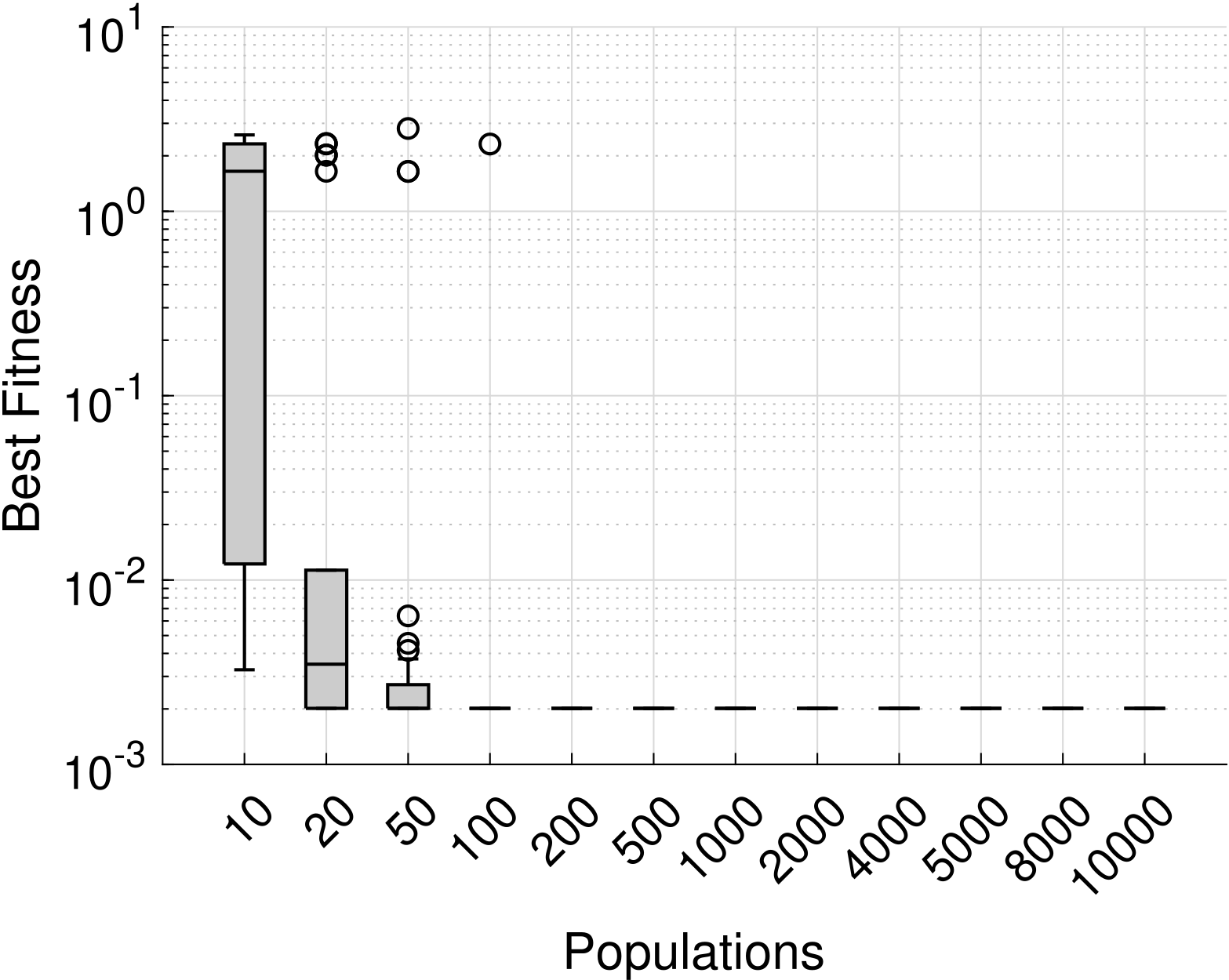}
        \caption{QIEO}
        \label{fig:AckleyPopvsBestFitQIEO}
    \end{subfigure}
    \caption{Ackley: Distribution of achieved fitness values across 30 trials for different population sizes.}
    \label{fig:AckleyfitnessVsPopulation}
\end{figure*}

\begin{figure*}[hbt]
    \centering
    \begin{subfigure}{0.47\textwidth}
        \centering
    \includegraphics[width=0.9\linewidth]{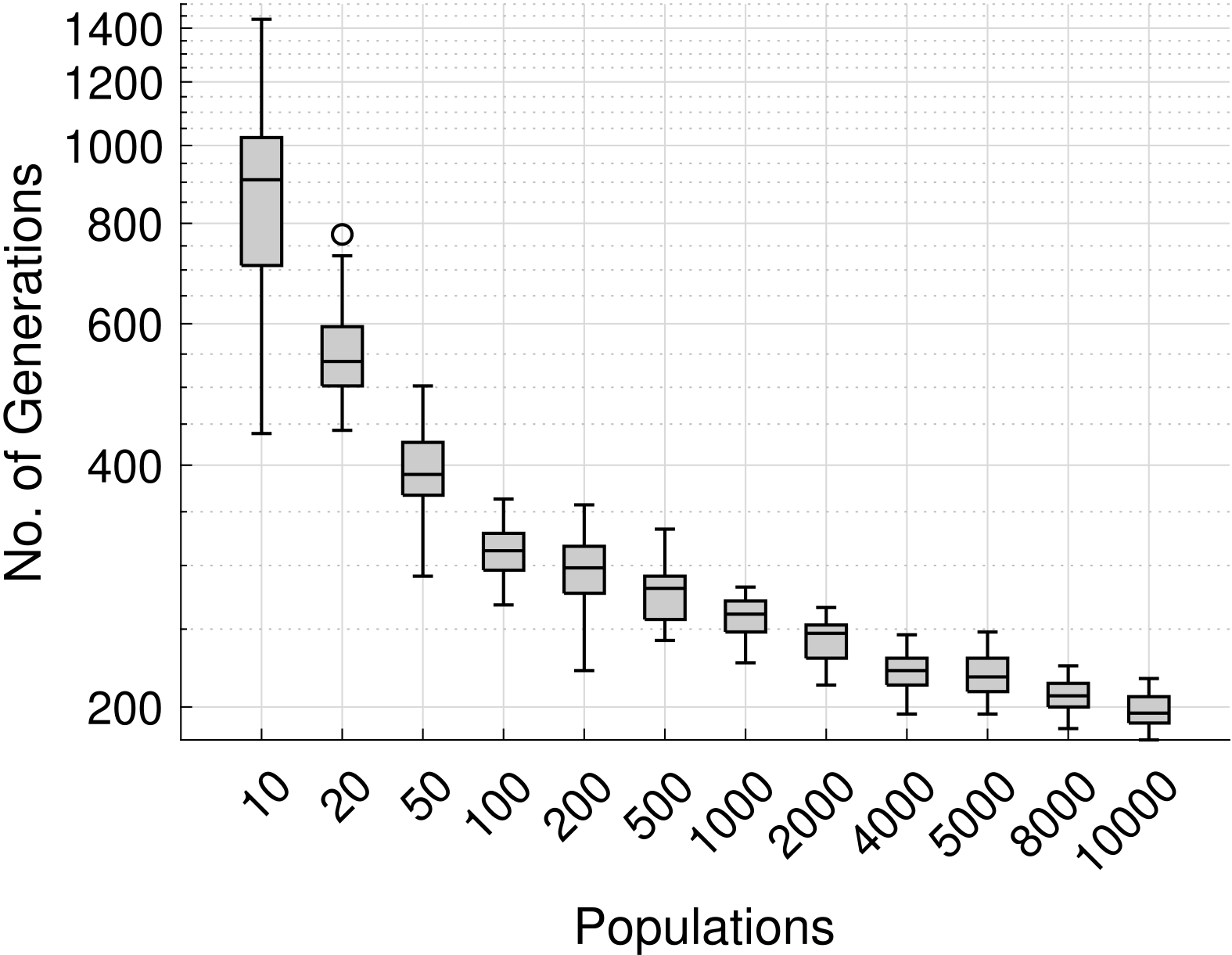}
    \caption{GA}
    \label{fig:AckelyPopvsGenGA}
    \end{subfigure}
\hfill
    \begin{subfigure}{0.47\textwidth}
        \centering
    \includegraphics[width=0.9\linewidth]{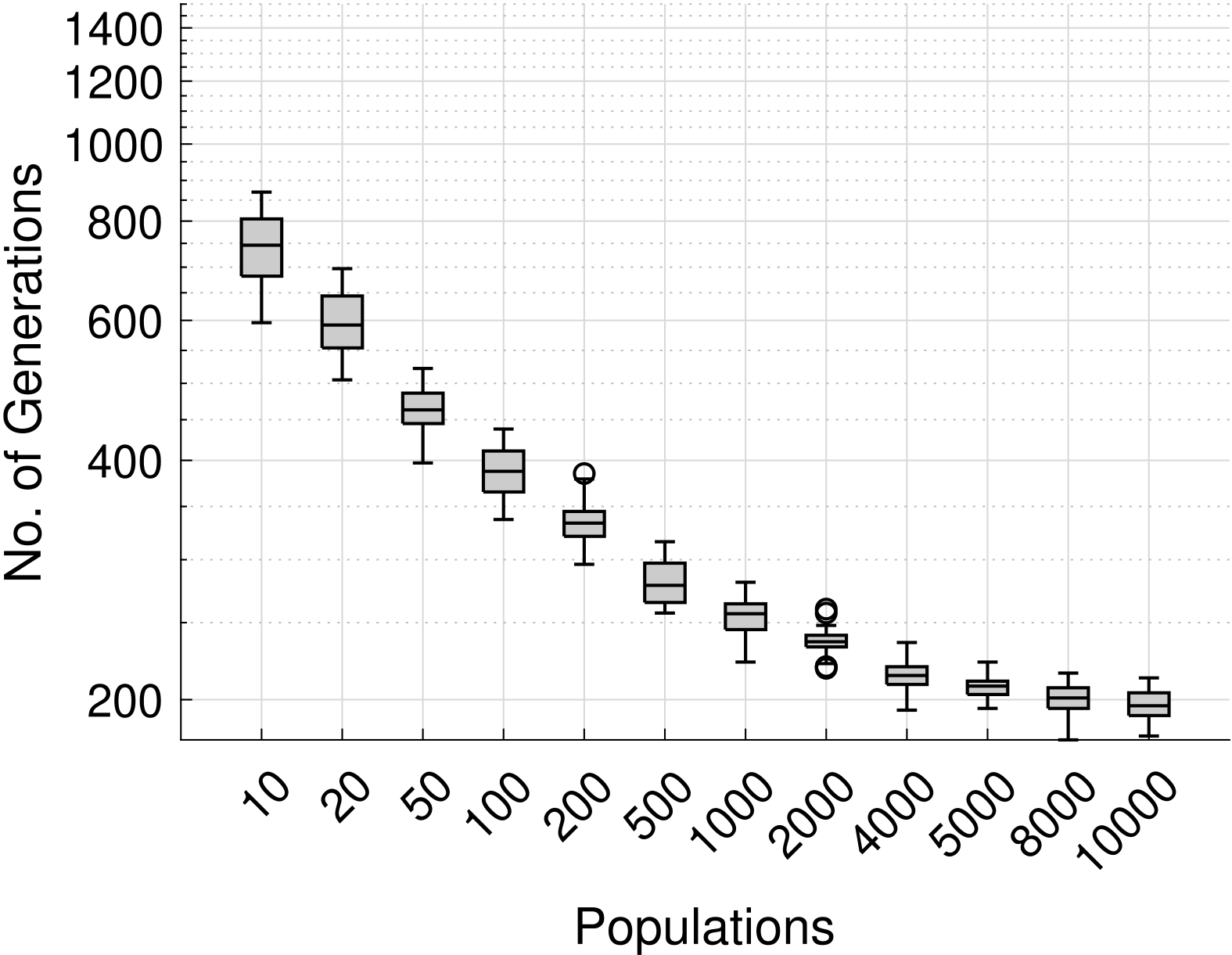}
    \caption{QIEO}
    \label{fig:AckelyPopvsGenQIEO}
    \end{subfigure}
\caption{Ackley: Distribution of convergence rates across 30 trials for different population sizes.}
\label{fig:AckleyPopvsGen}
\end{figure*}

\begin{figure*}[hbt]
    \centering
    \begin{subfigure}{0.47\textwidth}
        \centering
        \includegraphics[width=0.9\linewidth]{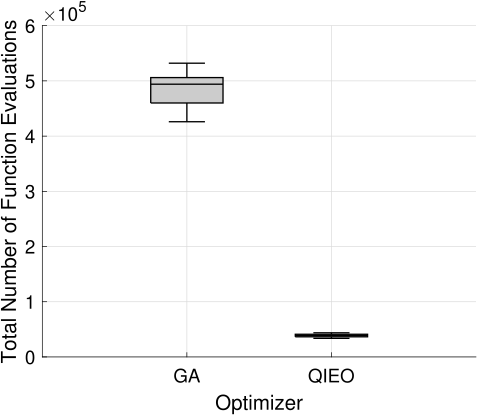}
        \caption{Function Evaluations for GA vs. QIEO}
        \label{fig:ackleyFuneval}
    \end{subfigure}
    \hfill
    \begin{subfigure}{0.47\textwidth}
        \centering
        \includegraphics[width=1.1\linewidth]{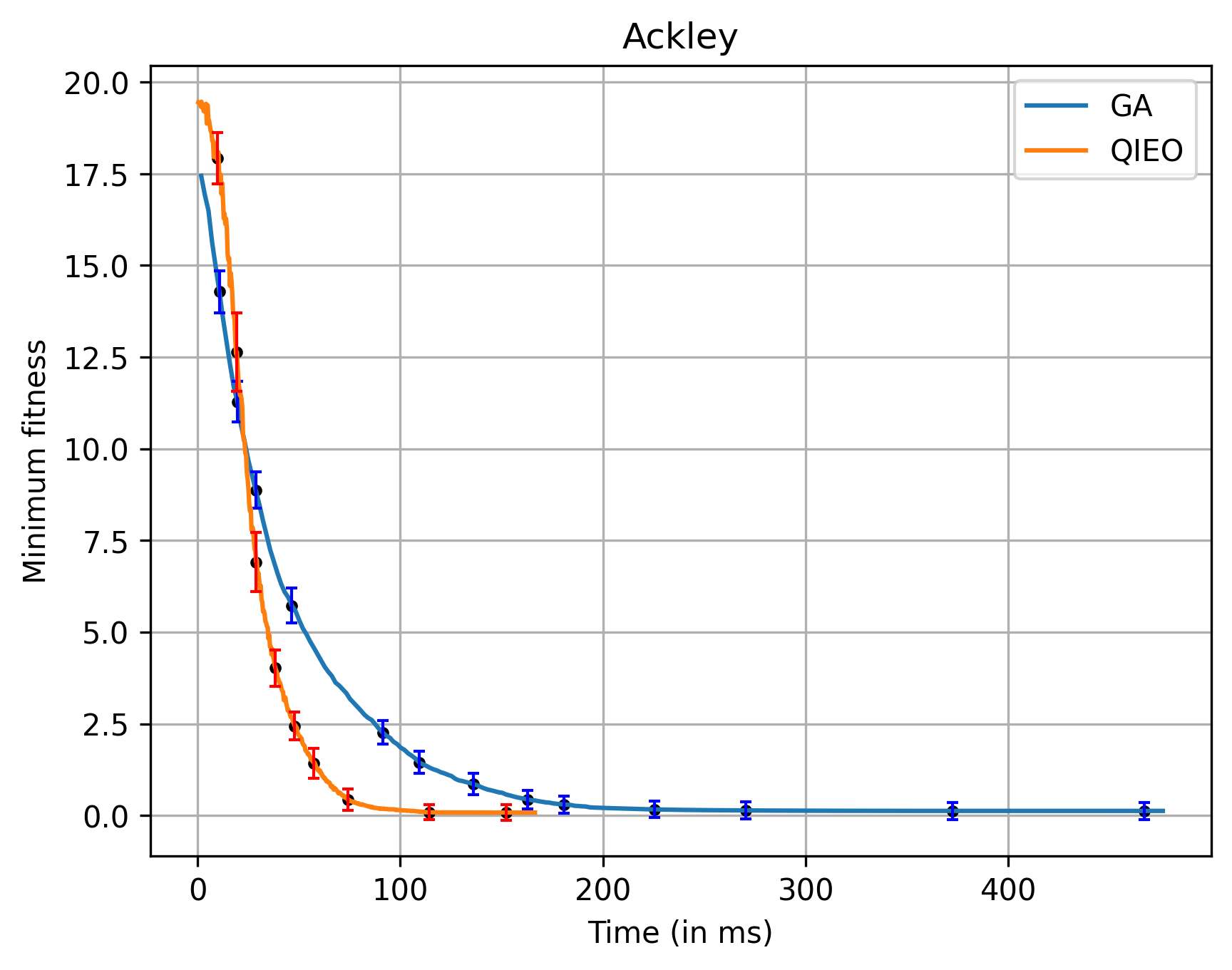}
        \caption{Convergence curves of GA and QIEO}
        \label{fig:ackleyConvergence}
    \end{subfigure}
    \label{fig:ackleyEvaluationsConvergence}
    \caption{Ackley: Comparison of GA and QIEO, showing
the distribution of the total number of function evaluations across 30 trials to achieve the same fitness and the evolution of the average fitness over time across 30 trials.}
\end{figure*}

\begin{figure*}[hbt]
    \centering
    \begin{subfigure}{0.47\textwidth}
        \centering
        \includegraphics[width=0.9\linewidth]{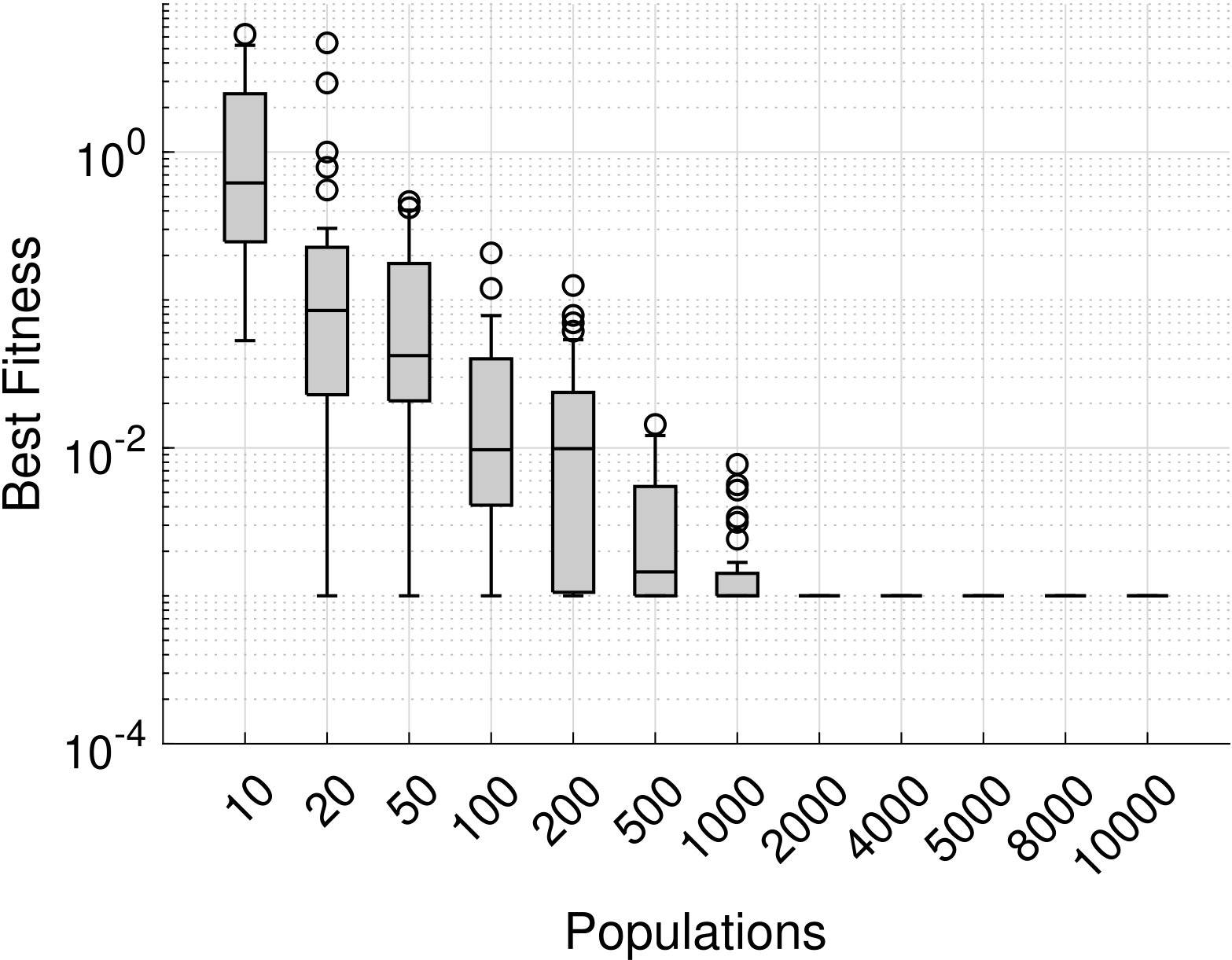}
        \caption{GA}
        \label{fig:RosenPopvsBestFitGA}
    \end{subfigure}
    \hfill
    \begin{subfigure}{0.47\textwidth}
        \centering
        \includegraphics[width=0.9\linewidth]{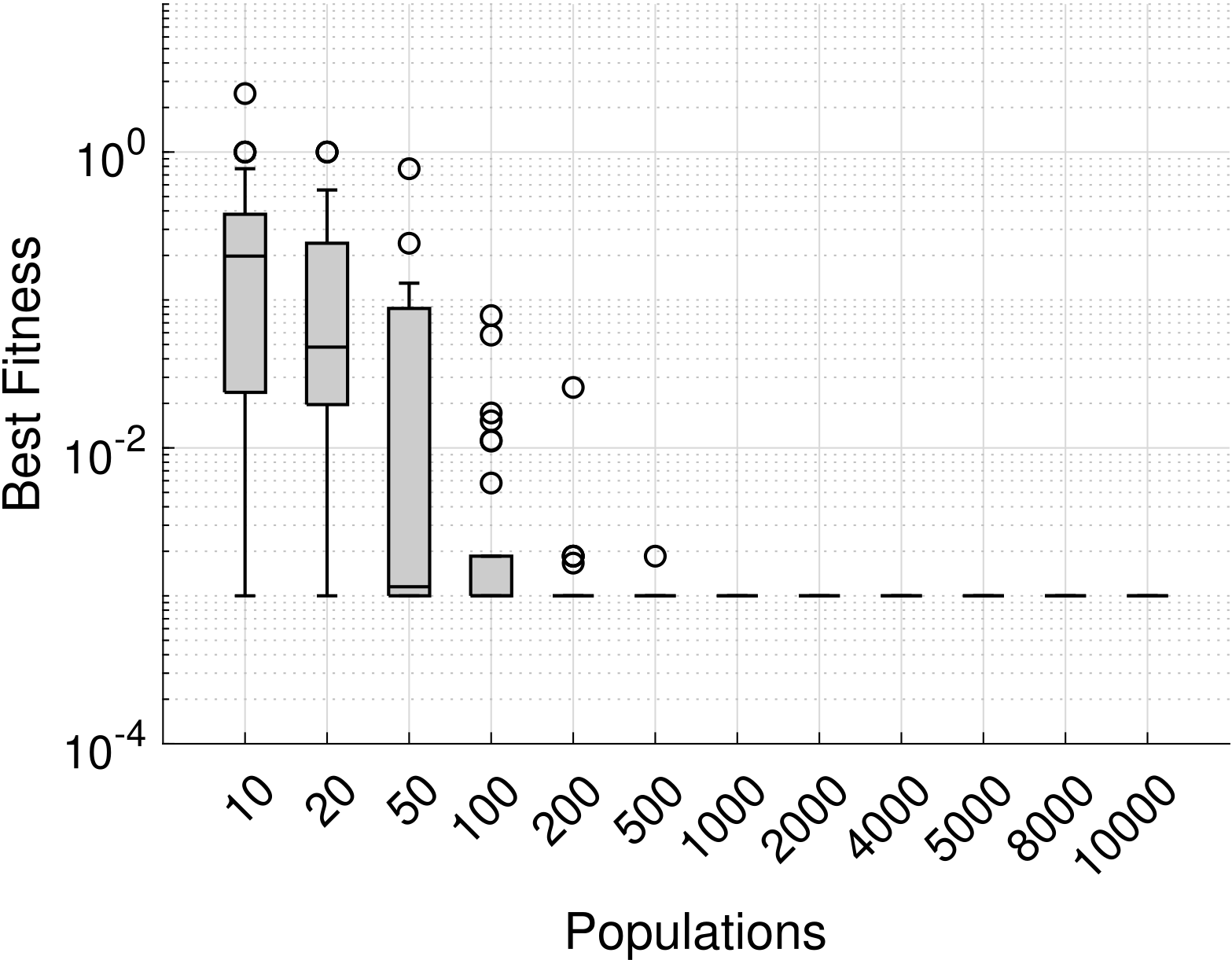}
        \caption{QIEO}
        \label{fig:RosenPopvsBestFitQIEO}
    \end{subfigure}
    \caption{Rosenbrock: Distribution of achieved fitness values across 30 trials for different population sizes.}
    \label{fig:RosenfitnessVsPopulation}
\end{figure*}

\begin{figure*}[hbt]
    \centering
    \begin{subfigure}{0.47\textwidth}
        \centering
    \includegraphics[width=0.9\linewidth]{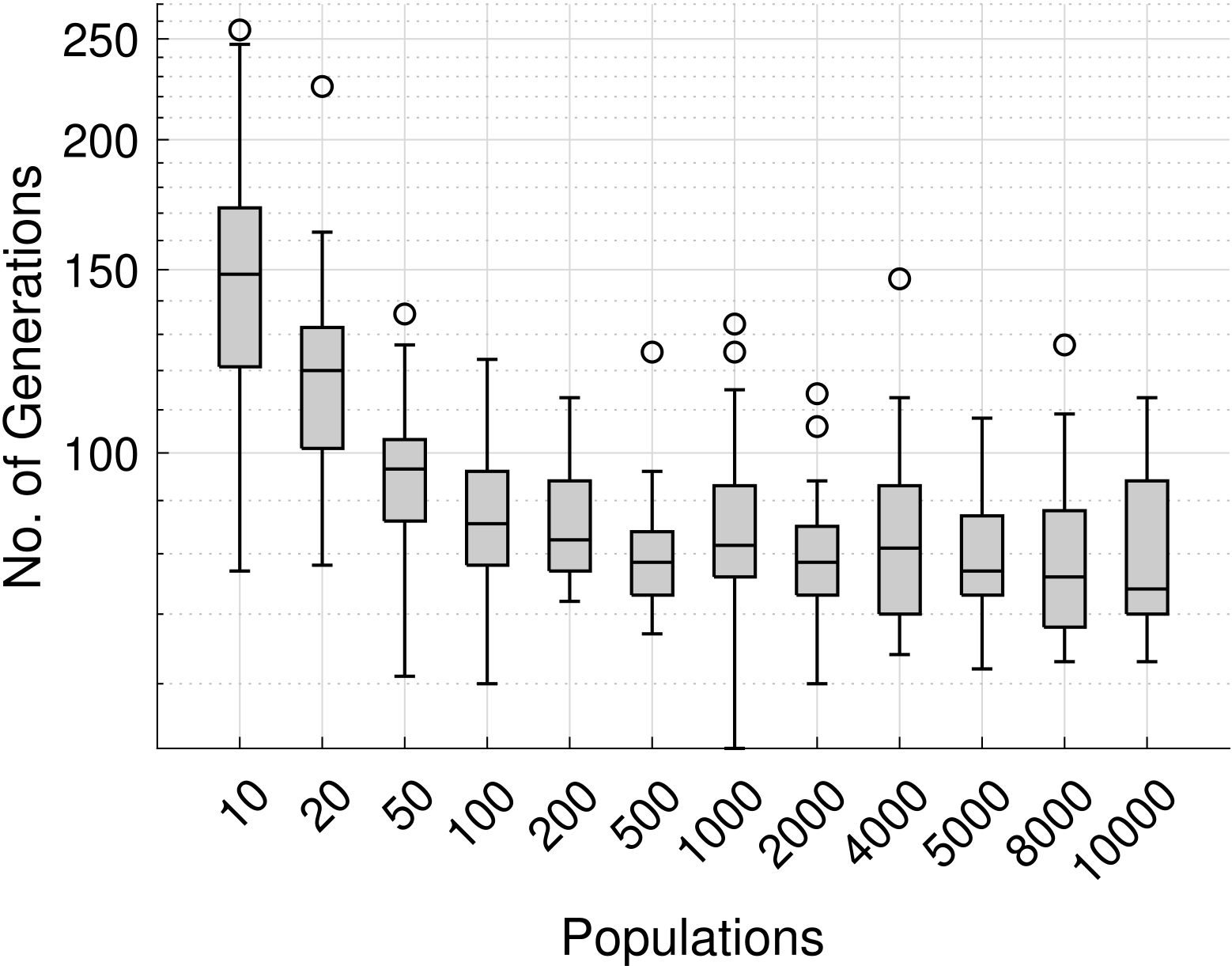}
    \caption{GA}
    \label{fig:RosenPopvsGenGA}
    \end{subfigure}
\hfill
    \begin{subfigure}{0.47\textwidth}
        \centering
    \includegraphics[width=0.9\linewidth]{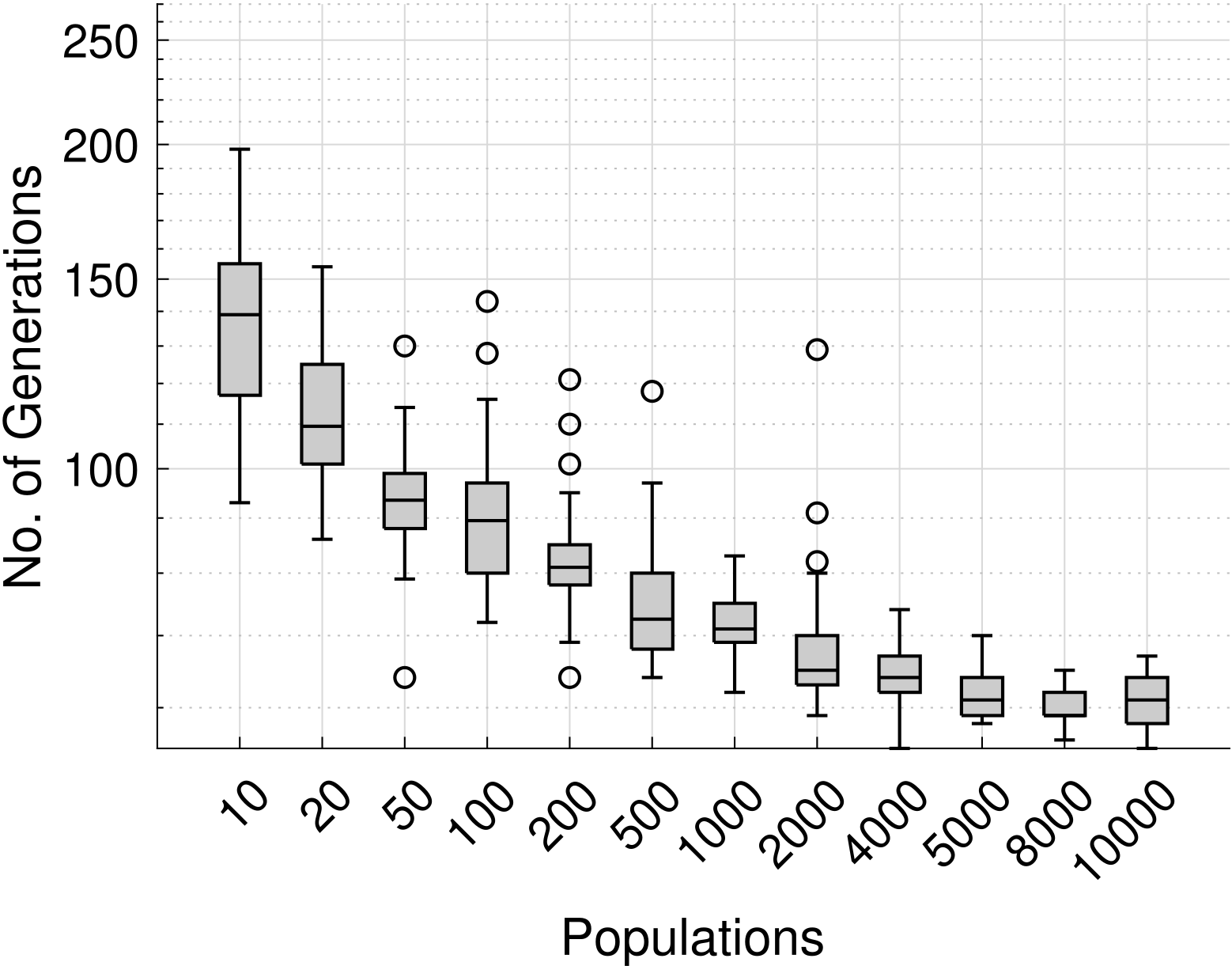}
    \caption{QIEO}
    \label{fig:RosenPopvsGenQIEO}
    \end{subfigure}
\caption{Rosenbrock: Distribution of convergence rates across 30 trials for different population sizes.}
\label{fig:RosenPopvsGen}
\end{figure*}

\begin{figure*}[hbt]
    \centering
    \begin{subfigure}{0.47\textwidth}
        \centering
        \includegraphics[width=0.9\linewidth]{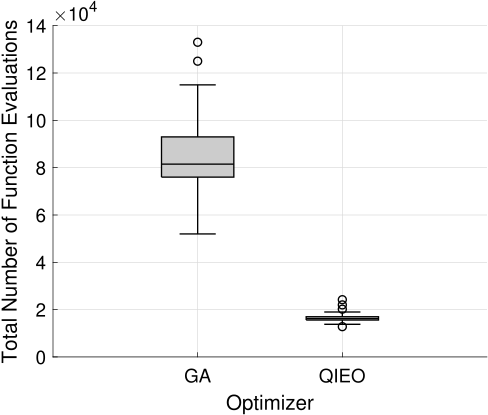}
        \caption{Rosenbrock: Function Evaluations}
        \label{fig:RosenFuneval}
    \end{subfigure}
    \hfill
    \begin{subfigure}{0.47\textwidth}
        \centering
        \includegraphics[width=1.1\linewidth]{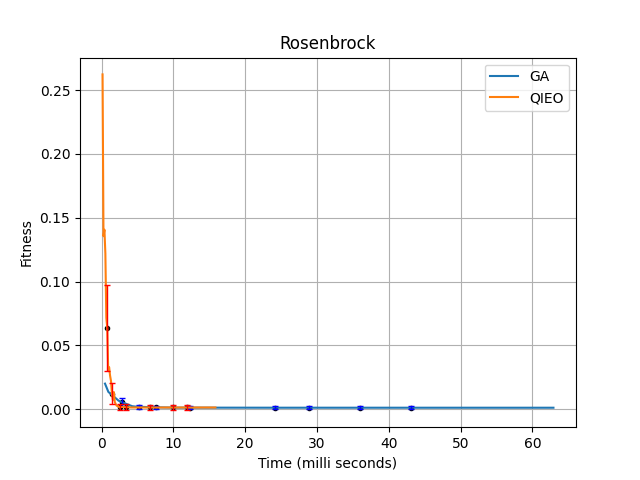}
        \caption{Rosenbrock: Convergence Plot}
        \label{fig:RosenConvergence}
    \end{subfigure}
\caption{Rosenbrock: Comparison of GA and QIEO, showing the distribution of total function evaluations across 30 trials to achieve the same fitness, and the evolution of average fitness with time}
    \label{fig:RosenEvaluationsConvergence}
\end{figure*}

\begin{figure*}[hbt]
    \centering
    \begin{subfigure}{0.47\textwidth}
        \centering
    \includegraphics[width=0.9\linewidth]{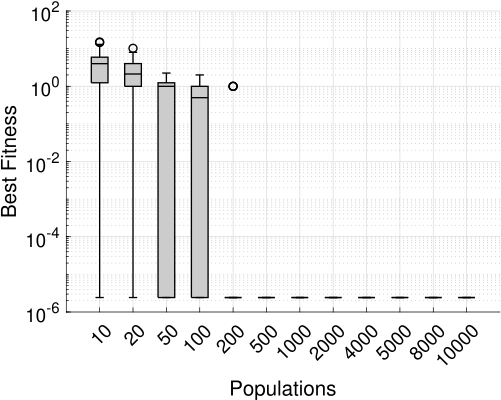}
    \caption{GA}
        \label{fig:RastriginPopvsBestFitGA}
    \end{subfigure}
    \hfill
    \begin{subfigure}{0.47\textwidth}
        \centering
        \includegraphics[width=0.9\linewidth]{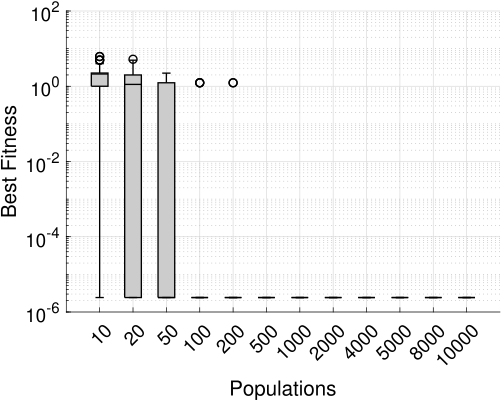}
        \caption{QIEO}
    \label{fig:rastriginPopvsBestFitQIEO}
    \end{subfigure}
    \caption{Rastrigin: Distribution of achieved fitness values across 30 trials for different population sizes.}
    \label{fig:rastriginfitnessVsPopulation}
\end{figure*}

\begin{figure*}[hbt]
    \centering
    \begin{subfigure}{0.47\textwidth}
        \centering
    \includegraphics[width=0.9\linewidth]{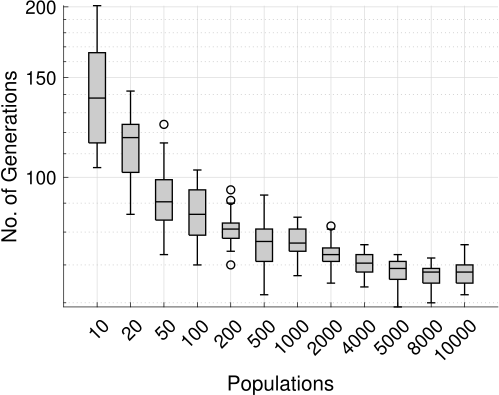}
    \caption{GA}
    \label{fig:rastriginPopvsGenGA}
    \end{subfigure}
\hfill
    \begin{subfigure}{0.47\textwidth}
        \centering
    \includegraphics[width=0.9\linewidth]{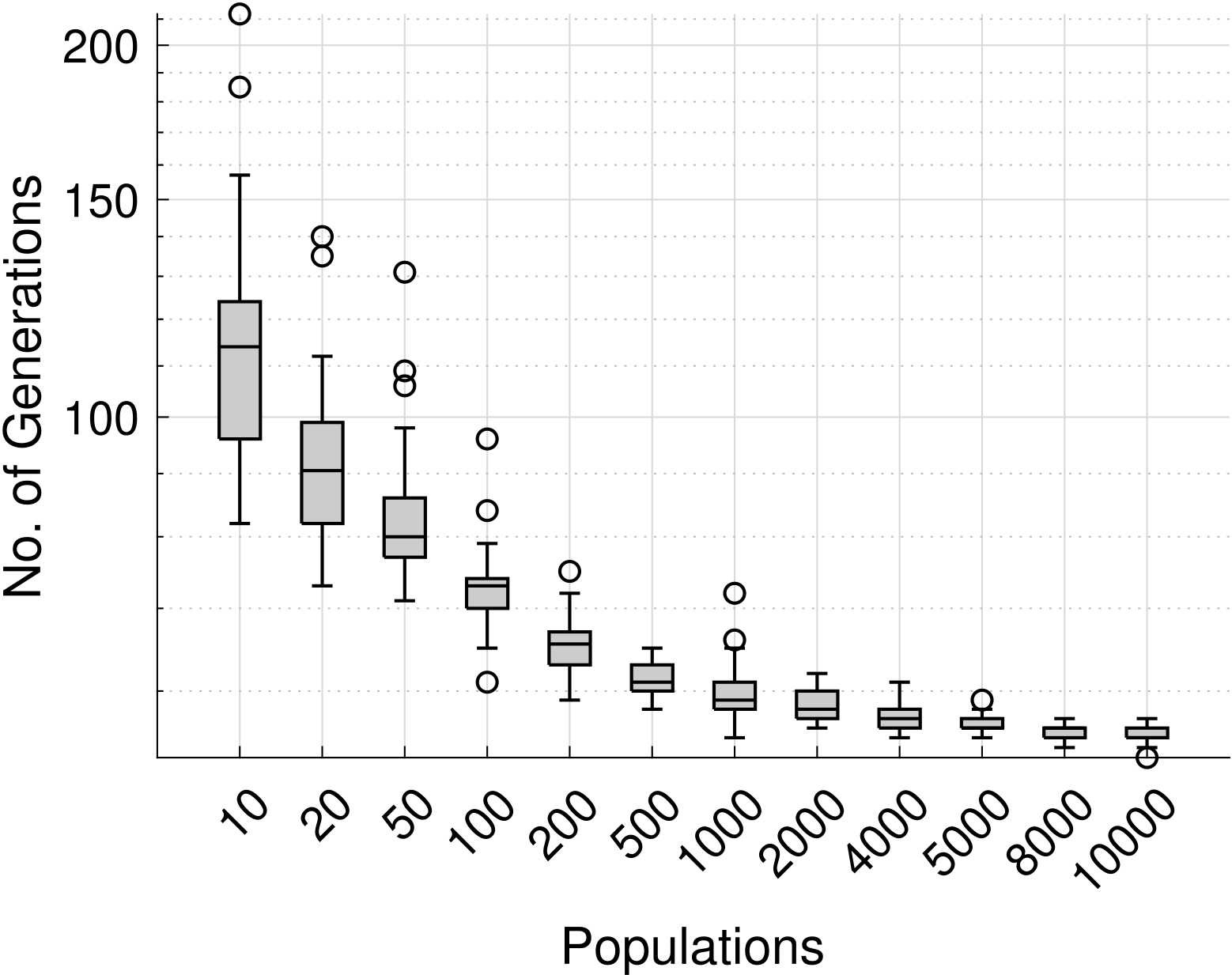}
    \caption{QIEO}
    \label{fig:rastriginPopvsGenQIEO}
    \end{subfigure}
\caption{Rastrigin: Distribution of convergence rates across 30 trials for different population sizes.}
\label{fig:rastriginPopvsGen}
\end{figure*}

\begin{figure*}[hbt]
    \centering
    \begin{subfigure}{0.47\textwidth}
        \centering
    \includegraphics[width=0.9\linewidth]{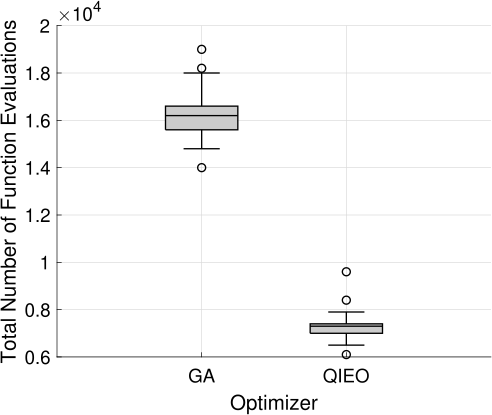}
    \caption{GA vs. QIEO comparison of total number of function evaluations}
    \label{fig:rastriginfuneval}
    \end{subfigure}
\hfill
    \begin{subfigure}{0.47\textwidth}
        \centering
    \includegraphics[width=1.1\linewidth]{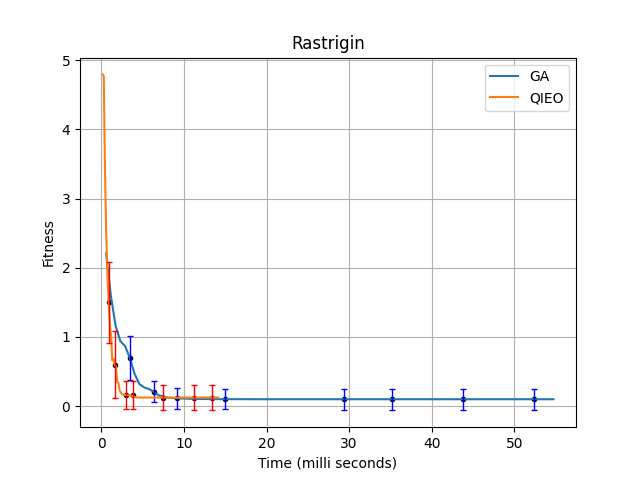}
    \caption{GA vs. QIEO: Convergence curve}
    \label{fig:rastriginconvergence}
    \end{subfigure}
\caption{Rastrigin: Comparison of GA and QIEO, showing the distribution of total function evaluations across 30 trials to achieve the same fitness, and the evolution of average fitness over time across 30 trials.}
\end{figure*}

\begin{figure*}[hbt]
    \centering
    \begin{subfigure}{0.47\textwidth}
        \centering
        \includegraphics[width=0.9\linewidth]{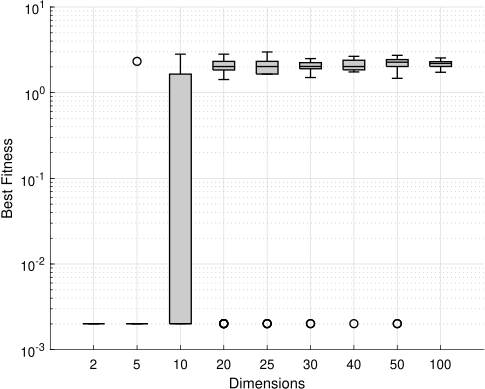}
        \caption{GA}
        \label{fig:AckelyDimvsFitGA}
    \end{subfigure}
    \hfill
    \begin{subfigure}{0.47\textwidth}
        \centering
        \includegraphics[width=0.9\linewidth]{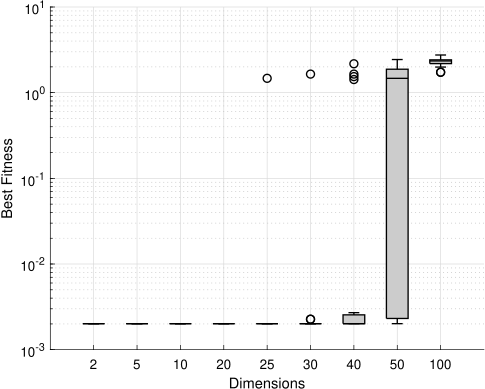}
        \caption{QIEO}
        \label{fig:AckelyDimvsFitQIEO}
    \end{subfigure}
\caption{Ackley: Distribution of achieved fitness across 30 trials for different dimensions.}
    \label{fig:Ackleydim}
\end{figure*}

\begin{table*}[hbt]
\centering
\renewcommand{\arraystretch}{1.5} % Adjust row height
\caption{GA vs. QIEO comparison with respect to total number of function evaluations to reach same accuracy for the three different functions}
\begin{tabular}{|c|p{1.5cm}|p{1.5cm}|p{1.5cm}|p{1.5cm}|p{1.5cm}|p{1.5cm}|}
%{|c|p|p|p|p|p|p|}
\hline
{Functions} & \multicolumn{3}{c|}{{GA}} & \multicolumn{3}{c|}{{QIEO}} \\ \hline
& {Population size} & {Generations} 
& {Function evaluations} & {Population size} & {Generations} & {Function evaluations} \\ \hline
Ackley         &   2000            &                  243            &   486000            &    100                           &    387            &   38700            \\ \hline
Rosenbrock         &  1000                             &     85      &  85000              &   200             &                    84             &    16800           \\ \hline
Rastrigin        &      200                          &    82            &   16400            &      100                           &        73        &    7300           \\ \hline
\end{tabular}
\label{tab:ga_qieo}
\end{table*}

\begin{table*}[hbt]
\centering
\caption{Speed-up factor comparison of GA and QIEO for the three different functions}
%\resizebox{\textwidth}{!}
{
\begin{tabular}{|l|l|l|l|}
\hline
{Functions} & {Time for GA (ms)} & {Time for QIEO (ms)} & {Speed up} \\ \hline
Ackley & 430.79 & 148.05 & 2.9 \\ \hline
Rosenbrock & 62.93 & 15.85 & 3.9  \\ \hline
Rastrigin &54.73 &14.15 & 3.84 \\ \hline
\end{tabular}
}
\label{tab:ga_qieo_speedup}
\end{table*}

The performance of both algorithms, QIEO and GA, is evaluated through a comprehensive comparison of the results obtained by varying their population sizes. The following two key metrics are used for an effective comparison:
\begin{itemize}
    \item Accuracy: The ability of the algorithm to reach the fitness value within the given tolerance. The three selected benchmark functions have the global minimum as zero. A solution that is close to zero is generally considered tolerance. The tolerance value for the Ackley and the Rosenbrock is $1e^{-3}$, while for the Rastrigin, it is $1e^{-6}$.
    \item Convergence rate: The number of generations required by the algorithm to reach the specified accuracy. 
\end{itemize}

To analyze the influence of population size on performance, the following populations were tested: 10, 20, 50, 100, 200, 500, 1000, 2000, 4000, 5000, 8000, and 10,000. Each optimization problem is evaluated across 30 trials to check the repeatability and reliability of the algorithm. The results are presented using box-and-whisker plots to provide a comprehensive view of performance variation. Figure \ref{fig:AckleyfitnessVsPopulation} shows the fitness variation of the Ackley function for different population sizes using the GA and QIEO. 
Similar plots for the Rosenbrock function are provided in Fig. \ref{fig:RosenfitnessVsPopulation}, and for the Rastrigin function in Fig. \ref{fig:rastriginfitnessVsPopulation}, %for GA and Fig. \ref{fig:rastriginPopvsBestFitQIEO} for QIEO),
illustrating the accuracy of both algorithms. From these accuracy plots (Figs. \ref{fig:AckleyfitnessVsPopulation}, \ref{fig:RosenfitnessVsPopulation}, and \ref{fig:rastriginfitnessVsPopulation}), it is evident that GA consistently requires larger population sizes than QIEO for satisfying the convergence criteria. For example, in the case of the Ackley function (ten-dimensional), GA requires a population size of 2000 to achieve the desired fitness accuracy in all trials. In contrast, QIEO accomplishes the same result with a population size of only 100. This disparity is smaller for the Rosenbrock and Rastrigin functions (two-dimensional); for Rosenbrock, GA needs approximately 1000 individuals, while QIEO requires 200. In the Rastrigin function, GA requires 200 individuals compared to QIEO's 100. Despite the smaller gap, QIEO consistently converges with smaller population sizes.

Figure \ref{fig:AckleyPopvsGen} presents the convergence rate variation of the Ackley function for different population sizes from the GA and the QIEO. 
The similar plots for Rosenbrock and Rastrigin are presented in Figs. \ref{fig:RosenPopvsGen} and \ref{fig:rastriginPopvsGen} respectively.
From these convergence rate plots, even when using the same population size for both QIEO and GA, in most cases, QIEO requires fewer generations to reach the desired accuracy. This is evident from the lower maximum points in the box-and-whisker plots for QIEO compared to GA across most population sizes. 
The time efficiency naturally follows from the average total number of function evaluations, with related plots shown in Figs. \ref{fig:ackleyConvergence}, \ref{fig:RosenConvergence}, and \ref{fig:rastriginconvergence}.
These plots visually represent the convergence process over time for GA and QIEO, averaged over 30 trials. The bars showcase the variance of the fitness value over 30 trials for that particular time. The population sizes used for demonstrating time efficiency are listed in Table \ref{tab:ga_qieo}, and it is clear from these plots that QIEO consistently converges faster than GA to the desired accuracy.

In order to analyze the impact of dimensionality on the optimizer, we vary the dimension of the problem by keeping the population size constant for both GA and QIEO. An increase in dimension inherently increases the number of design variables/unknowns of the optimization problems and provides insights into the scalability of the algorithm. 
The Ackley function is chosen for this exploration due to its complex nonlinear design space, and the following dimensions are chosen: 2, 5, 10, 20, 25, 30, 40, 50, 100. The variation of fitness across 30 trials is presented in Fig. \ref{fig:Ackleydim}. From the dimension plots, we observe that QIEO finds the desired fitness in all the trials across most dimensions, with the exception of the higher ones -- 50 and 100. In stark contrast, GA fails to approach the desired fitness in almost all dimensions and exhibits a wide variance, performing poorly overall and only achieving the desired fitness with low variance only in the lower dimensions -- 2 and 5. 

If an algorithm achieves the desired fitness with a smaller population size, it might require more generations to do so, which diminishes any practical advantage due to the increased number of function evaluations. 
Therefore, to accurately assess an algorithm’s efficiency, we must consider the average total number of function evaluations required to achieve the target fitness. 
This number can be derived by multiplying the population size by the average convergence rate across 30 trials derived from the convergence rate plots. For instance, in the Ackley function, QIEO requires a population size of 100 to reach the desired accuracy. By multiplying this population size with the average convergence rate of 387, we calculate a total of 38,700 function evaluations. In contrast, GA requires a population size of 2000 to achieve the same fitness, with an average convergence rate of 243 generations. This results in 486,000 evaluations, indicating that GA demands 12 times more evaluations than QIEO to reach the same fitness level. Similar calculations for the Rastrigin and Rosenbrock functions reveal that GA requires 5.1 and 2.2 times more function evaluations than QIEO. The average convergence rate for the desired accuracy, the population size required, and the total number of function evaluations required for all three functions are tabulated in Table \ref{tab:ga_qieo}. The distribution of the total number of function evaluations to achieve the desired fitness across 30 trials is illustrated in Figs. \ref{fig:ackleyFuneval}, \ref{fig:RosenFuneval}, \ref{fig:rastriginfuneval}. It is evident that QIEO exhibits less variance compared to GA, making it a more reliable choice.

The speedup achieved by QIEO is calculated as the ratio of the average time taken by GA to the average time taken by QIEO across 30 trials to reach the desired accuracy. It provides insights on the 
GA takes 430.79 milliseconds for the Ackley function, while QIEO takes 148 milliseconds, making QIEO 2.9 times faster. Similarly, for the Rosenbrock and Rastrigin functions, QIEO is faster by a factor of $3.9$ and $3.84$, respectively, as detailed in Table \ref{tab:ga_qieo_speedup}.

We have established that QIEO performs well with smaller population sizes and achieves the desired accuracy in fewer generations. This is due to the enhanced exploration of the overall search space provided by QIEO, which helps limit premature convergence \cite{han2006quantum}. Now, consider a scenario in GA where we start with a very small population, say 10 individuals. If these individuals begin in a poor region of the search space, for them to explore different regions, the crossover point—chosen randomly—must occur in the significant bits of the individual. Suppose each design variable is represented by 16 bits; the likelihood of making substantial changes to the variable depends on the probability of the crossover point being selected in these significant bits, as well as the number of crossover events per generation. 
With fewer individuals, there are fewer crossover opportunities, which reduces diversity in the offspring created \cite{grefenstette1987incorporating}. While the mutation operator could change the significant bits, mutation rates are typically kept low. This means the probability of altering the most significant bits is also low, limiting the algorithm's ability to escape unpromising regions of the search space. So, the significant bits may not undergo many changes, thereby forcing the algorithm to stay in the unpromising regions of the search space. This could be one way in which the exploration process is hindered, leading to premature convergence in GA. 
There's another common reason that, irrespective of the encoding of why GA struggles with smaller population sizes, random events like which individuals are selected for reproduction can lead to certain genes becoming more common by chance, not because they are the best solutions. Over time, this can cause the algorithm to "drift" towards certain solutions, even if they are not optimal, a phenomenon known as genetic drift \cite{10089573} \cite{797972}. 

In contrast, in QIEO, even with a small population starting in a poor region from the initial measurement of the uniform superposition, the application of the $Ry$ gate with a small angle ensures that the most significant qubits have an equal likelihood of changing as any other qubits. This enables QIEO to have the most significant qubits with similar probabilities that can be collapsed to $1$ and $0$. 
The rotation gate only partially aligns the most significant qubit with the best individual’s significant bit, thereby promoting exploration through the most significant bits. As a result, QIEO enhances exploration and reduces the likelihood of premature convergence even with lower population sizes \cite{1134125}. 
The global exploration capabilities of QIEO, without the deterioration of local search capabilities as a result of the probabilistic measurement process, seem to provide additional potential for convergence to better solutions. 

The variance in convergence rate plots
(Figs.~\ref{fig:ackleyConvergence}, \ref{fig:RosenConvergence}, \ref{fig:rastriginconvergence}) for QIEO is consistently lower than that of GA across all cases. Both algorithms, GA and QIEO, initiate the search process randomly, meaning each trial starts from a different set of points. Ideally, an efficient algorithm should require roughly the same number of generations to achieve the desired accuracy, regardless of the starting points, minimizing its dependence on them. QIEO’s lower variance suggests it excels in this regard. This can be attributed to the fact that, at the start, the probability amplitudes of the qubits in quantum individuals are close to $\frac{1}{\sqrt{2}}$. As a result, the quantum individuals have a high probability of collapsing into various regions of the search space, facilitating an effective global search. In contrast, GA appears to be more reliant on the initial population. For example, if GA starts in a region of the search space with few optimal minima, the tournament selection operator will still choose the best individuals from this population, but those individuals may still be far from optimal because the region lacks high-quality solutions. This makes GA more dependent on the initial population distribution, contributing to its higher variance. When the crossover and mutation operators are applied to this population, they will recombine and slightly modify these already sub-optimal solutions. While crossover might introduce some new combinations, it is essentially working with poor-quality building blocks \cite{Whitley1994}, making it less likely to create highly fit individuals. Mutation, though capable of introducing diversity, tends to introduce changes, which may sometimes not be sufficient to break out of the low-quality region of the search space. As a result, GA can get stuck exploring unpromising areas of the search space for several generations before eventually finding better regions, or it might converge prematurely on local optima. This dependency on initial conditions often leads to higher variance in GA's convergence rate across different trials. This can also lead to premature convergence when compared to QIEO, which is better equipped to explore the entire search space early on due to the idea inspired by the uniform quantum superposition. In essence, GA’s reliance on selection, crossover, and mutation can sometimes hinder its ability to efficiently navigate the search space if it starts in a poor region or operates with small population sizes. At the same time, QIEO’s quantum characteristics allow it to explore more globally from the outset, reducing its dependence on initial conditions or population size. The quantum nature of the individuals allows for a thorough exploration of the entire search space early on. The algorithm gradually gains a good understanding of the landscape and transitions steadily into a more focused search, zeroing in on the region containing the high-quality solutions. This local search phase happens after a few generations in the QIEO \cite{1134125}.

A key challenge in classical meta-heuristic algorithms is determining suitable parameter values. Navigating the design space effectively while iteratively improving solution optimality requires a balance between local and global search capabilities. 
However, achieving this balance is difficult, as prioritizing one often diminishes the other. In genetic algorithms (GA), this balance is managed by multiple operators—primarily crossover and mutation—both of which are parameter-dependent. Understanding how these parameters affect performance and tuning them to optimize results for a specific problem is a complex task \cite{10.1145/3377929.3398136}.  On the other hand, QIEO simplifies this process with only one parameter \emph{---} Ry gate, paired with the measurement operation. This combination naturally balances exploitation and exploration. Moreover, both components are well-suited for parallelization, enabling independent evolution of individuals without the need for complex parameter tuning \cite{10.1145/2598394.2598437}. This streamlined approach reduces the challenge of parameter optimization, making QIEO an effective and scalable solution for complex optimization problems.

Achieving a suitable balance between exploration, which entails global search, and exploitation, which involves local search, has remained a persistent challenge in meta-heuristic algorithms. Quantum computing concepts in QIEO have bolstered the global search capability without deteriorating the local search capabilities \cite{1688636, Hakemi2024}. A key feature of QIEO is the updating of quantum individuals based on the best solution found in previous iterations. By adjusting the quantum individuals’ probability amplitudes using the $R_Y$ gate, they are guided to have a slightly higher likelihood of collapsing near the best solution. This adjustment helps preserve solution quality while maintaining diversity. The small probabilities associated with this process ensure that quantum individuals do not fully converge to the best solution, leaving a significant chance for them to collapse into other regions of the search space, thereby promoting exploration. The ability of quantum individuals to exist in a superposition of solutions—both near the best solution and in other parts of the search space—enables a balanced search behavior. When measurement occurs, some quantum individuals collapse near the best solution (exploitation), while others collapse into different parts of the search space (exploration). This mechanism replaces traditional operators like crossover and mutation, which are typically used in evolutionary algorithms to improve solutions and maintain diversity. By leveraging quantum principles, QIEO effectively balances guided behavior toward the best solutions and exploration of new areas in the search space. 

\section{Conclusions}
This study evaluates the performance of the GPU-optimized QIEO algorithm in comparison to the GPU-optimized GA on three benchmark functions: Ackley, Rastrigin, and Rosenbrock. The results clearly show that QIEO significantly outperforms GA, requiring far fewer function evaluations to achieve the desired fitness —- 12 times fewer for Ackley, 5 times fewer for Rosenbrock, and 2.5 times fewer for Rastrigin. 
This reduction in function evaluations translates to a substantial decrease in time taken to convergence, with Ackley completing it in one-third of the time and Rosenbrock and Rastrigin in one-fourth of the time compared to GA. 
Additionally, QIEO demonstrates lower variance in both fitness outcomes and convergence rates across 30 trials, underscoring its superior reliability. 
This combination of enhanced reliability, reduced computational effort, and faster optimization makes QIEO particularly valuable for engineering optimization problems, where the function evaluation is the costliest process. 

\section*{Acknowledgment}
The authors would like to thank the Ministry of Heavy Industries (Government of India) and the Indian School of Business, who supported K.E.S.K, S.B.S., and A.A. for the duration of this project. The authors would like to thank the team of the Center for Computational Research at the University of Buffalo, which gave us the resources to support our work, both in computing capabilities and knowledge from their staff.
\bibliographystyle{unsrt}
\bibliography{Cite}

\end{document}